\documentclass[%
 reprint,longbibliography,preprintnumbers,
nofootinbib,
 amsmath,amssymb,
 aps,
pre,
]{revtex4-2}
\pdfoutput=1
\usepackage{graphicx,float}
\usepackage[utf8]{inputenc}
\usepackage{flushend}
\usepackage{dcolumn}
\usepackage{bm}
\usepackage{balance}

\usepackage{xcolor} 
\usepackage[normalem]{ulem} %

\usepackage[normalem]{ulem}

\usepackage{verbatim}
\usepackage{color,ulem}
\usepackage[english]{babel}
\newcommand\Wang[1]{\textcolor{blue}{ [W:\,#1]}}
\newcommand\Sun[1]{\textcolor{purple}{ [S:\,#1]}}

\def\be{\begin{equation}}
\def\ee{\end{equation}}
\def\bea{\begin{eqnarray}}
\def\eea{\end{eqnarray}}

\usepackage[utf8]{inputenc}

\input Starburst.fd
\newcommand*\initfamily{\usefont{U}{Starburst}{xl}{n}}\initfamily

\makeatletter 
    
\renewcommand\onecolumngrid{
\do@columngrid{one}{\@ne}%
\def\set@footnotewidth{\onecolumngrid}
\def\footnoterule{\kern-6pt\hrule width 1.5in\kern6pt}%
}

\renewcommand\twocolumngrid{
        \def\footnoterule{
        \dimen@\skip\footins\divide\dimen@\thr@@
        \kern-\dimen@\hrule width.5in\kern\dimen@}
        \do@columngrid{mlt}{\tw@}
}%

\makeatother

\newcommand{\beq}{\begin{eqnarray}}
\newcommand{\eeq}{\end{eqnarray}}
\usepackage{amsmath}
\usepackage{tikz}
\usetikzlibrary{decorations.pathmorphing}
\usetikzlibrary{shapes.misc}
\tikzset{cross/.style={cross out, draw=black, minimum size=8*(#1-\pgflinewidth), inner sep=0pt, outer sep=0pt},
cross/.default={1pt}}
\usetikzlibrary{patterns,math}
\usepackage[colorlinks = true,
            linkcolor = blue,
            urlcolor  = blue,
            citecolor =red,
            anchorcolor = blue]{hyperref}
            
\allowdisplaybreaks[4]

\def\arXiv#1{\href{http://arxiv.org/abs/#1}{arXiv:#1}}
\def\arXiv#1#2{\href{http://arxiv.org/abs/#1}{arXiv:#1}}

\def\doi#1#2{\href{http://doi.org/#1}{#2}}

\begin{document}

\title{\boldmath Quantum scars  
from holographic boson stars}
\vskip 0.5cm

\author{Yan Liu$^{1}$}
\email{yanliu@buaa.edu.cn}
\author{Ya-Wen Sun$^{2}$}
\email{yawen.sun@ucas.ac.cn}
\author{Yuan-Tai Wang$^{3,4}$}
\email{wangyuantai@ustc.edu.cn}

\affiliation{
\vspace{0.25cm}
$^1$ 
Department of Space Science and Peng Huanwu Collaborative Center for Research and Education, 
Beihang University, Beijing 100191, China}
\affiliation{$^2$School of Physical Sciences, University of Chinese Academy of Sciences, Beijing 100190, China}

\affiliation{$^3$Interdisciplinary Center for Theoretical Study, University of Science and Technology of China, Hefei 230026, China}
\affiliation{$^4$Peng Huanwu Center for Fundamental Theory, Hefei 230026, China}\vspace{-8pt}


\begin{abstract}
Quantum many-body scars are atypical nonthermal states embedded in the chaotic spectrum that evade conventional ergodicity. We propose the asymptotically AdS mini-boson star as a holographic candidate for scar-like states. Their 
spectrum exhibits random-matrix signatures of chaos while supporting embedded integrable spectral branches. The 
whole holographic system, including black holes, is generically chaotic with most eigenstates satisfying the eigenstate thermalization hypothesis; in contrast, the boson star macrostate probes 
a near-integrable subsector within this chaotic spectrum, signaling scarred spectral structures. Boson stars further display anomalously low entanglement relative to black holes at the same energy density, and also robust revivals in Krylov complexity, revealing nonergodic dynamics. These spectral, entanglement, and dynamical diagnostics provide unified evidence for holographic quantum scars in a self-gravitating system. Our work suggests a new connection between many-body scar physics, quantum chaos, and horizonless gravitational dynamics.
\end{abstract}

\preprint{USTC-ICTS/PCFT-26-26}

\maketitle

{\bf \textit{ Introduction.--}}
Understanding how isolated quantum many-body systems thermalize, and how this process can fail, is a central problem among quantum chaos, statistical mechanics, and gravity. A remarkable exception to conventional ergodicity is provided by quantum many-body scars, namely the atypical nonthermal states embedded within the otherwise chaotic spectrum that exhibit hidden integrable structures, suppressed entanglement, and anomalous revivals. Originally discovered in constrained quantum systems, the scar phenomena have since been identified across a growing range of models, suggesting that they arise from more general underlying principles rather than being tied exclusively to their original microscopic constraints 
\cite{Turner:2017fxc, Bernien:2017ubn, Serbyn:2020wys, Moudgalya:2021xlu}. 

At the same time, holographic many-body systems provide a natural arena to explore thermalization and chaos in strongly coupled quantum matter \cite{Zaanen:2015oix,Hartnoll:2016apf}. Black holes correspond to paradigmatic maximally chaotic thermal states \cite{Shenker:2013pqa, Maldacena:2015waa, Cotler:2016fpe} that obey the eigenstate thermalization hypothesis (ETH), while horizonless geometries may encode nonthermal sectors inaccessible within conventional black hole physics. This distinction raises a fundamental question: can quantum scars admit a gravitational or holographic realization? 

While C-stars were suggested as a possible candidate \cite{Milekhin:2023was}, a sharp spectral and dynamical realization of many-body scars in a self-gravitating holographic system remains absent. In this work, we propose that asymptotically AdS mini-boson stars provide a more compelling holographic realization of quantum scar-like states, supported by unified evidence from the spectral structure, entanglement behavior, and the revival dynamics. This proposal further points to a broader perspective: generic horizonless star geometries may realize scar-like states in holographic many-body systems. 
 
Boson stars are horizonless compact stellar objects formed by self-gravitating Bose-Einstein condensates of bosons \cite{Liebling:2012fv} that can serve as macroscopic quantum states in holographic systems. Unlike black holes, these horizonless boson stars carry zero thermal entropy, 
which underscores their nonthermal character and supports the existence of scar-like holographic states. 
By analyzing their linear normal modes, we uncover a spectral structure consisting of globally chaotic correlations coexisting with embedded integrable subsectors, distinguished from the holographic spectra of integrable vacuum AdS and maximally chaotic black holes. This coexistence provides spectral signatures characteristic of scarred eigenstates embedded in the chaotic spectra. 
We further establish independent information-theoretic and dynamical evidence for this proposal. Using holographic entanglement entropy, we show that boson stars exhibit anomalously low, subthermal entanglement compared to black holes at the same energy density. Using holographic Krylov complexity, we uncover robust revival dynamics absent in black holes. 

Together, chaotic spectral statistics, embedded integrable sectors, suppressed entanglement, and complexity revivals provide a unified set of signatures for holographic quantum scars.
Our results suggest that nonergodic scarred structures can emerge naturally in gravitational dynamics, establishing a bridge between quantum chaos, holography, and horizonless spacetimes.

\vspace{0.2cm}
{\bf \textit{ Setup and background solutions.}}-- We focus on the simplest horizonless solution: the spherically symmetric static AdS mini-boson star in 3+1 dimensions \cite{Liebling:2012fv, Buchel:2013uba}. We consider Einstein gravity minimally coupled to a complex scalar field, with the action 
\begin{equation}
\begin{split}
\begin{aligned}
    S &= \int d^4x\sqrt{-g} \,\left[ \frac{1}{16\pi G}\big(R-2\Lambda\big) + \mathcal{L}_m \right],\\
    \mathcal{L}_m &= -\partial_\mu\Phi\partial^\mu\Phi^* - U(|\Phi|),\quad U(|\Phi|) = m^2 |\Phi|^2,
\end{aligned}
\label{eq:action}
\end{split}
\end{equation}
where the cosmological constant $\Lambda = -3/L^2$. We set the AdS radius $L=1$ and $8\pi G=1$ henceforth. We choose $m^2 = -2$ without loss of generality, corresponding to a dual scalar operator of conformal dimension $\Delta_+ =3$. 

For stationary backgrounds, we adopt the standard spherically symmetric ansatz with a harmonic time dependence for the scalar,  
\begin{align}\label{background fields}
    ds^2 &= -A(r)^2N(r) dt^2 + \frac{1}{N(r)} dr^2 + r^2 \big(d\theta^2 + \sin{\theta}^2 d\varphi^2\big),\nonumber \\
    \Phi &= e^{-i\omega t} \phi(r)\,,
\end{align}
where $\omega > 0$ is the frequency of the condensate, and $A(r)$, $N(r)$, $\phi(r)$ are real radial functions. We impose the sourceless boundary condition for $\phi(r)$.
Pure AdS$_4$ is recovered with $A(r)=1,\;N(r)=r^2+1,\;\phi(r)=0$. 
The background equations and boundary conditions are summarized in Appendix \ref{sec S2}. 

\begin{figure}[h!]
\centering
\includegraphics[width=0.475\textwidth]{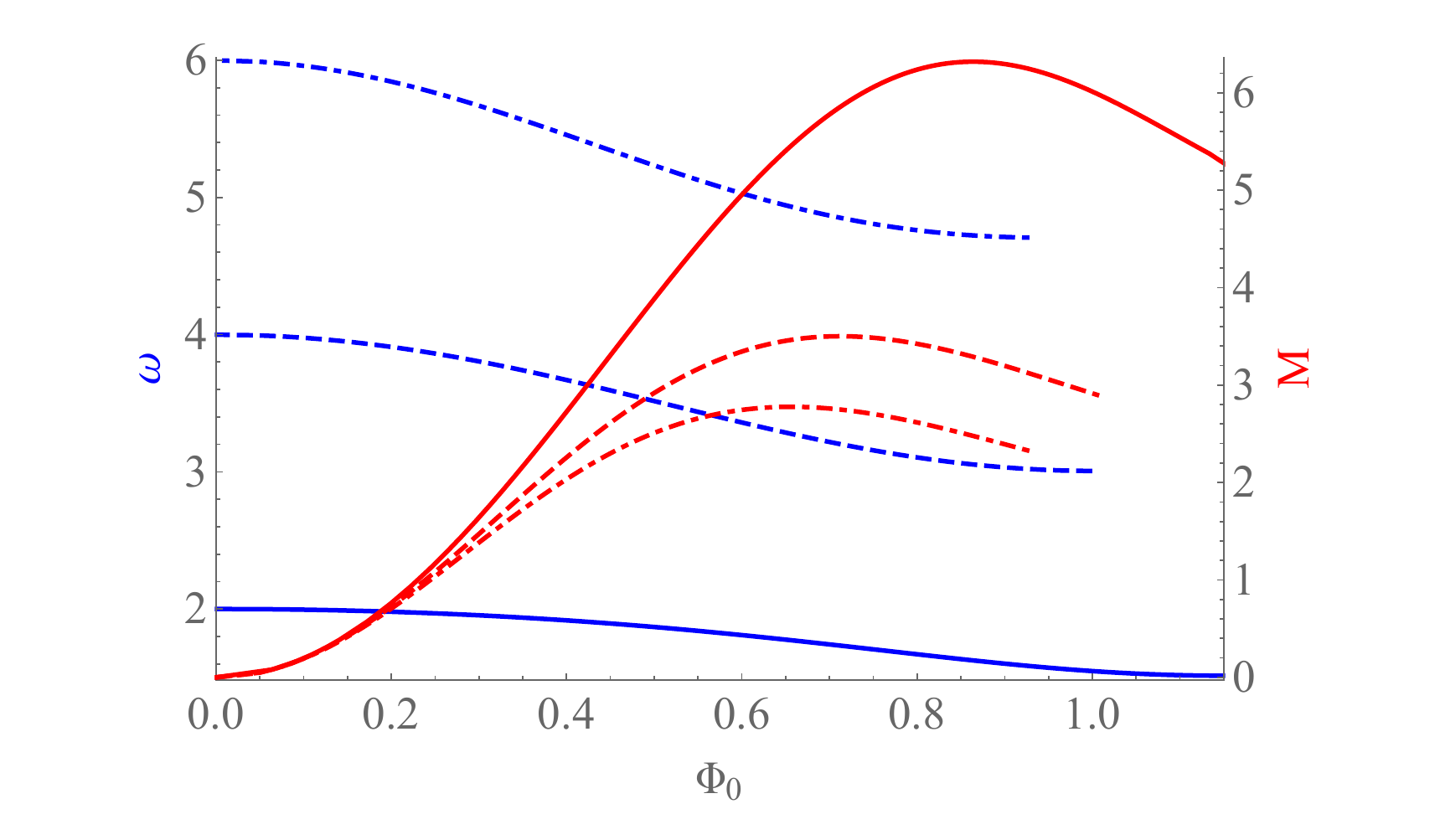}
\caption{Boson star frequency $\omega$ ({\em blue}) and mass $M$ ({\em red}) as functions of the central scalar amplitude $\Phi_0$. The solid, dashed, and dash-dot curves denote the ground state, first excited state, and second excited state, respectively. 
}
\label{fig_background}
\vspace{-6pt}
\end{figure}

Using a shooting method, we obtain one-parameter families of regular mini-boson star solutions for the ground state and the first few excited states. In principle, the construction can be extended to arbitrarily higher excited star states. 
All solution families are labeled by the central scalar amplitude $\Phi_0$, 
with the shooting parameter $A_0$ (the central value of $A$) and a bounded frequency $\omega$ determined dynamically.  
For each state, the mass $M$ has a maximal value and depends non-monotonically on $\Phi_0$, whereas $\omega$ decreases monotonically as $\Phi_0$ increases. 
These families of solutions interpolate between the AdS vacuum and heavy boson star configurations with large $M$ and $\Phi_0$, as
in Fig. \ref{fig_background}. A critical amplitude $\Phi_0^c$ where $M$ reaches its maximum marks the onset of instability: solutions beyond this point lie on the unstable branch and are expected to evolve toward gravitational collapse \cite{Liebling:2012fv,Khlopov:1985fch}. 
Representative field profiles of the second excited stars are shown in Fig. \ref{fig:profilebg} in Appendix \ref{sec S2}.

\vspace{0.2cm}
{\bf \textit{Linear perturbative spectrum.}}-- To probe the stability and spectral properties of these backgrounds, which are crucial for identifying scarred versus thermal behavior, we perform a linear perturbation analysis around each boson star background. We consider even-parity perturbations and focus on $\ell=0$ (spherically symmetric) and $\ell \geq 2$ (quadrupole and higher) sectors.
For the $\ell=0$ sector, the linearized metric and scalar perturbations take the form
\begin{equation}
\begin{split}
\begin{aligned}
    &\delta ds^2
    = e^{-i\Omega t}\big(-A(r)^2 N(r) H_0(r) dt^2 + \frac{H_2(r)}{N(r)} dr^2 \big) + c.c.,\\
    &\delta\Phi
    = e^{-i\omega t} \big(e^{-i\Omega t}\delta\phi_+(r) + e^{i\Omega t}\delta\phi_-(r)\big)\,,
 \end{aligned}
 \label{eq:fluc-ell0}
 \end{split}
 \end{equation}
 with real normal mode frequency $\Omega$, while for $\ell\geq 2$ 
 \begin{equation}
 \begin{split}
 \begin{aligned}    
   &\delta ds^2
   = e^{-i\Omega t} \big(-A(r)^2N(r) H_0(r)dt^2 - 2i\Omega r H_1(r) dtdr \\
   &+ \frac{H_2(r)}{N(r)} dr^2 + r^2 K(r)(d\theta^2 + \sin{\theta}^2  d\varphi^2)\big)Y_{lm}(\theta,\varphi) + c.c.,\\
    &\delta\Phi
    = e^{-i\omega t} \big(e^{-i\Omega t}\delta\phi_+(r)Y_{lm}(\theta,\varphi) + e^{i\Omega t} \delta\phi_-(r) Y_{lm}^*(\theta,\varphi)\big)\,.
\end{aligned}
\label{eq:fluc-ell2}
\end{split}
\end{equation}  
{For $\ell\geq 2$, the linearized equations imply $H_0(r) = -H_2(r)$.} Full perturbation equations and boundary conditions are given in Appendix \ref{sec S2}. 

{We solve them both analytically in the WKB approximation (high-frequency limit) and numerically.} For large $\Omega$, the normal-mode spectrum admits a WKB analysis. We write collectively all radial perturbations as 
\begin{equation}
    \psi(r) = e^{\Omega S(r)} \Big(\psi^{(0)}(r)+\frac{\psi^{(1)}(r)}{\Omega}+\frac{\psi^{(2)}(r)}{\Omega^2}+\cdots\Big)\,,
    \label{eq:WKB-ansatz}
\end{equation} 
where $S(r)$ is the eikonal phase function. 
For $\ell=0$, the scalar sector yields two asymptotic branches quantized by the sourceless boundary condition,
\begin{equation}
\label{l=0 modes}  
\Omega_n = \frac{n+1/2}{I_1}\pi\mp\omega,\qquad n = 0,1,2,\;\dots\; ,
\end{equation}
where 
$I_1 = \int_\epsilon^{r_{UV}} \frac{1}{A(r)N(r)}dr$. In the AdS vacuum they reduce to the evenly spaced probe scalar spectrum.

For $\ell\geq 2$, an additional independent gravitational degree of freedom produces a third branch,
\begin{equation}
\label{l2 modes revised}
\Omega_n=
\frac{n+1/2}{I_1}\pi\,,\qquad n = 0,1,2,\;\dots\; ,
\end{equation}
which reduces to the AdS gravitational spectrum in the vacuum limit. Thus, at large frequency the spectrum organizes into two interleaving branches for $\ell=0$, and three for $\ell\geq 2$. 

{To obtain the full normal-mode spectrum beyond the high-frequency WKB regime, we solve the linearized equations numerically using a shooting method.} For generic $\Omega$, the coupled gravitational-scalar normal modes 
are computed numerically from the vanishing 
of the source determinant 
\begin{equation}
\label{numeric modes}
    \text{Det }\mathcal{S}(\Omega) = 0\,,
\end{equation}
where $\mathcal{S}(\Omega)$ is the 
UV source matrix constructed from three independent solutions. 

\begin{figure}[h!]
  \centering
   \includegraphics[width=0.22\textwidth]{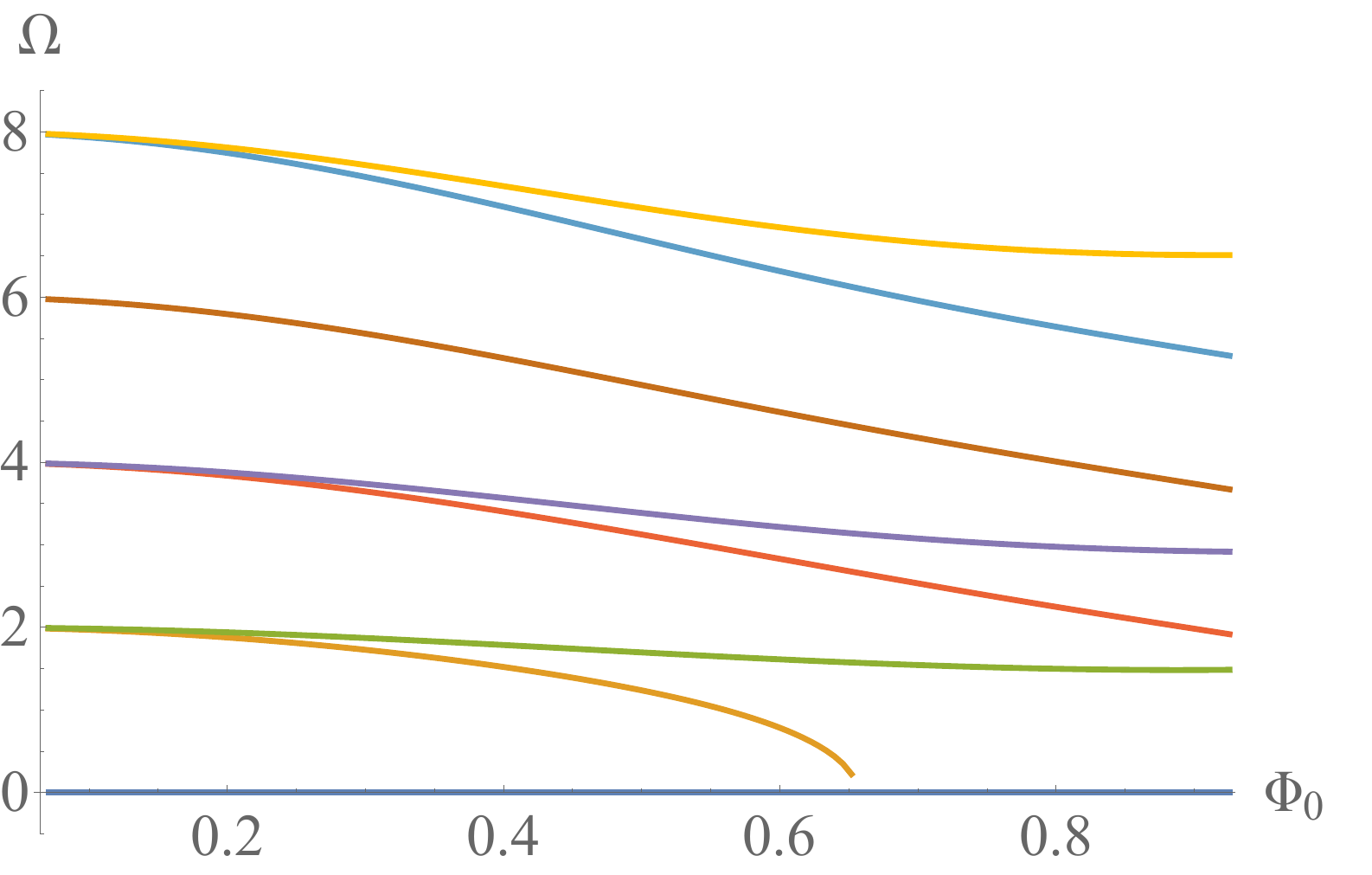}
    \includegraphics[width=0.21\textwidth]{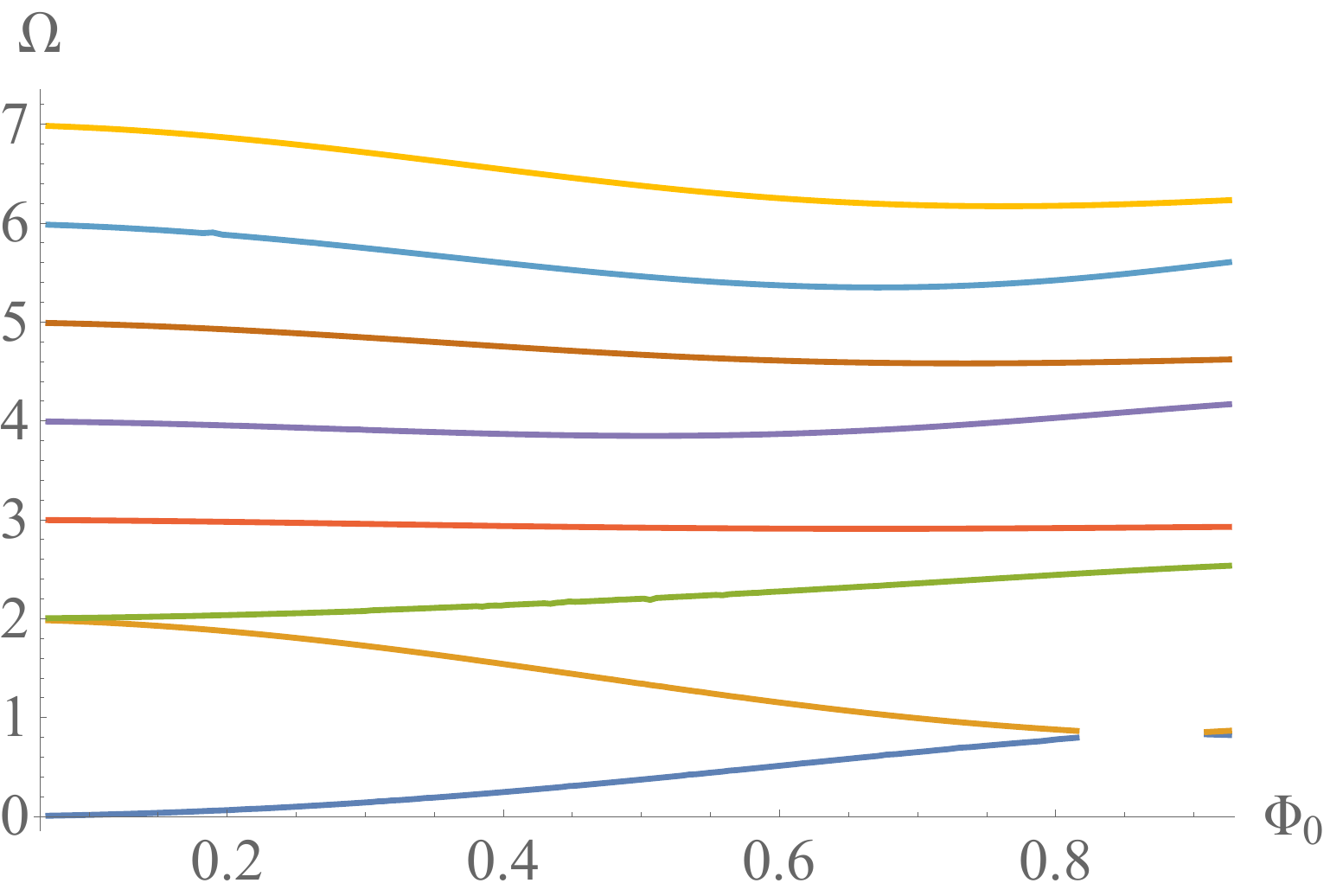} 
  \caption{\label{fig_modes}The 
  second excited-star normal modes as a function of $\Phi_0$ for $\ell=0$ ({\em left}) and $\ell=2$ ({\em right}).
  From bottom to top, the curves correspond to $n=1$-st$,\cdots, 8$-th modes. 
  }
\vspace{-6pt}
\end{figure}

Representative spectra for $\omega=\omega_{\text{min}}\simeq 4.708$ and $\omega=5$ 
for the second excited star
 are shown in Appendix \ref{sec S3}. The numerical spectra reproduce the interleaving branch structure predicted by WKB, with level spacings in excellent agreement up to expected subleading WKB corrections.
Across the full boson-star solution space, the modes interpolate continuously between vacuum AdS modes and heavy-star excitations, as shown in Fig.~\ref{fig_modes} 
for the second excited star as an example. 
Although gravitational and scalar perturbations are coupled away from vacuum, all modes can be continuously traced back to decoupled AdS
modes in the vacuum
limit \cite{Natario:2004jd,Dias:2013sdc}.


A notable feature is that vacuum degeneracies are rapidly lifted away from pure AdS. In particular, certain special modes cease to admit real roots of $\text{det}\,\mathcal{S}(\Omega)=0$ once $\Phi_0>\Phi_0^c$, signaling avoided crossings of modes. {Their analytic continuation into the complex plane develops a negative imaginary part, indicating linear instability}, consistent with the critical threshold discussed above.


\vspace{0.2cm}
{\bf \textit{RMT analysis.--}} With the normal-mode spectrum in hand, 
we probe chaos and hidden integrable structures using random matrix theory (RMT), treating the boson-star normal modes as an effective excitation spectrum around a nonvacuum holographic state. In the dual description, these frequencies precisely correspond to the normal modes  
and encode dynamical spectral information of the boundary theory.

\begin{figure}[h!]
\centering
\includegraphics[width=0.4\textwidth]{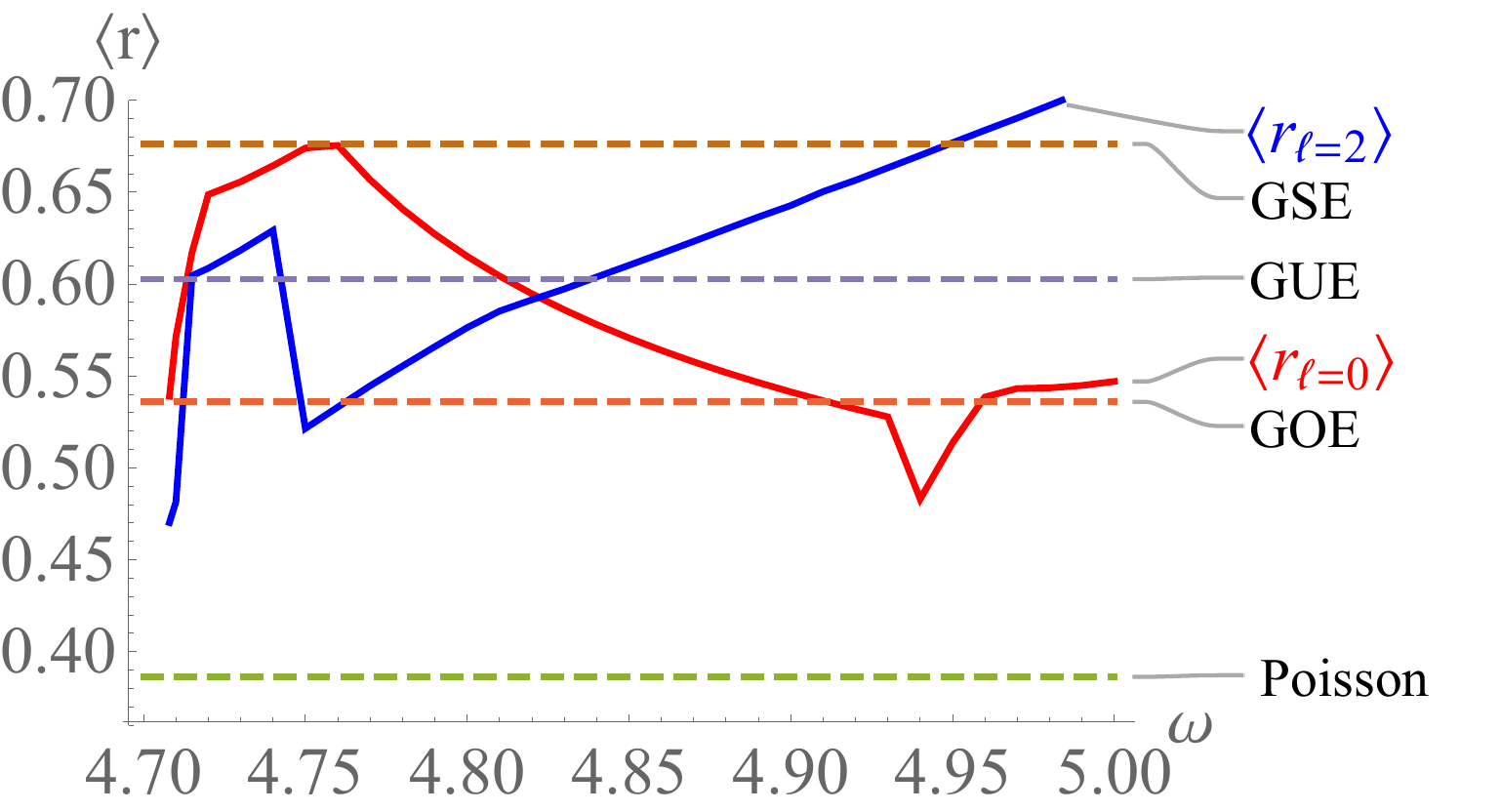}
\caption{\label{AGR}Average gap ratio of low-lying modes for the $\ell=0$ ({\em red}) and $\ell=2$ ({\em blue}) sectors of the second excited star. Dashed lines denote the universal RMT values for Poisson, Gaussian orthogonal, unitary, and symplectic ensembles (from bottom to top).
}
\vspace{-6pt}
\end{figure}

At small $n$, the spectrum displays an irregular distribution, as shown in Fig. \ref{fig_modes}, while at large $n$, the spectrum exhibits nearly equal level spacings. Therefore, to isolate genuine spectral correlations from this branch structure, we evaluate the average gap ratio $\langle r\rangle$ defined in \eqref{eq:gapratio}, which characterizes the degree of neighboring energy level repulsion, using the uncontaminated low-lying modes for each background solution (see Appendix~\ref{sec S1}).
The result for the second excited star is shown in Fig.~\ref{AGR}. 



For heavy excited stars, the average gap ratio $\langle r \rangle$ interpolates between standard RMT universality classes in both $\ell=0$ and $\ell=2$ sectors, falling squarely in the random matrix regime, i.e. Gaussian orthogonal ensemble $ \text{(GOE)}\approx 0.536$ to Gaussian symplectic ensemble $\text{(GSE)}\approx 0.676$. This distribution reveals robust chaotic level correlations, indicating that the generic fluctuations around the scarred macrostate still feel the underlying chaotic nature of the holographic CFT.

At large $n$,
however, the modes decouple into distinct branches, each exhibiting nearly equal level spacings. This decoupling arises because, in the large-$n$ limit, each branch is dominated by a single source. The Dirichlet boundary condition together with the AdS asymptotics enforce an approximately equally spaced tower. 
As a result, the chaotic mixing between the branches is suppressed, revealing a clear integrable subsector. 
Compared to the spectrum of the second excited star, the ground-state and first excited star's normal modes can be studied in a similar way, which start to gather into the equally spaced tower at a much smaller $n$, exhibiting stronger integrability.

Within the spectrum of the boson star, the coexistence of chaotic statistics at low $n$ and integrable equal-spacing structure at high $n$ provides a sharp spectral signature of 
quantum many-body scars: 
the whole holographic system, including both black holes and horizonless stars as distinct boundary macrostates, remains chaotic overall, with most eigenstates obeying ETH; by contrast, the particular macrostate dual to the boson star resides within and unveils a near-integrable subsector embedded in this otherwise chaotic spectrum.
This observation is
consistent with a weak form of ETH violation  
on the boundary and 
supports a scar-like interpretation of the holographic stars.  



\vspace{0.3cm}
{\bf \textit{Lower entanglement entropy.--}} Scarred states in chaotic quantum systems are distinguished by anomalously low entanglement relative to thermal states at the same energy density. We show that boson stars exhibit precisely such subthermal entanglement signatures compared to maximally chaotic black holes.

We study the holographic entanglement entropy using Ryu–Takayanagi (RT) formula \cite{Ryu:2006bv} 
\begin{equation}\label{RT}
    S_A = \frac{\text{Area}(\Gamma_A)}{4G} + \cdots,
\end{equation}
where the RT surface $\Gamma_A$ for the boundary region $A$ is the minimum bulk
surface homologous to the boundary region
$A$. We pick $A$ to be
a hemisphere and define $\delta S_{half}$ as its vacuum-subtracted entanglement entropy, which is a finite quantity. For the background metric~\eqref{background fields}, the corresponding area functional \eqref{RT} is
\begin{equation}
    \text{Area}(\Gamma_A)
    =4\pi\int_{
    r_{min}}^\infty r\sin{\theta(r)}\sqrt{\frac{1}{N(r)}+r^2\theta'(r)^2}\;dr\,,
\end{equation}
where $r_{min}$ is the radial turning point of the RT surface parameterized by $\theta(r)$ which is determined by extremization.

\begin{figure}[h!]
\centering
\includegraphics[width=0.4\textwidth]{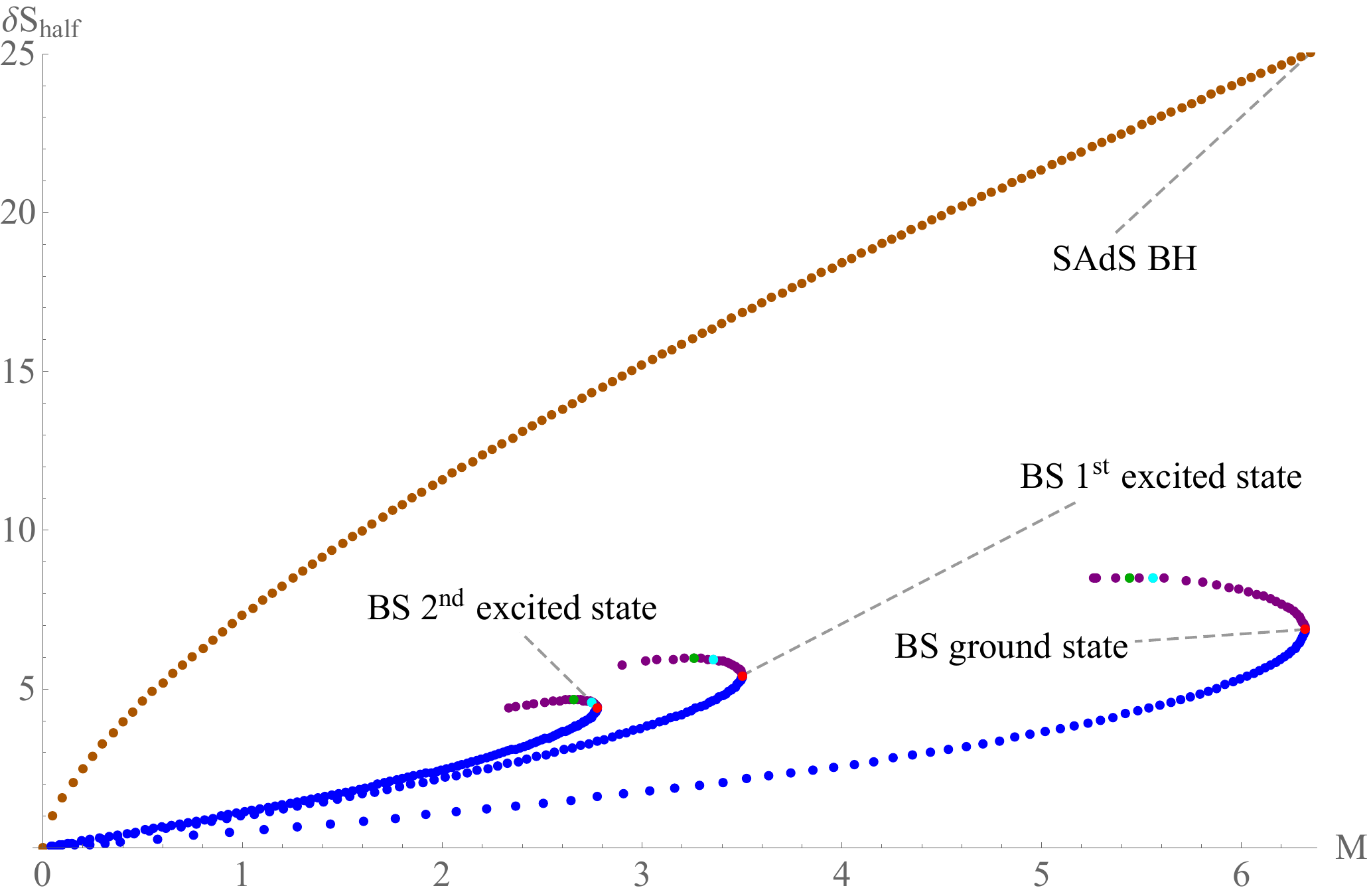}
\caption{\label{S(M)} The vacuum-subtracted entanglement entropy $\delta S_\text{half}$ as a function of the mass $M$
in SAdS black hole ({\em upper} brown dotted curve) and the ground-state, first, and second excited-state boson star (three {\em lower} blue and purple dotted curves). The critical points on
each boson star curve correspond to maximum $\delta S_{half}$ ({\em green}), the transition of RT surface configurations ({\em cyan}), and maximum $M$ ({\em red}), respectively.
 }
\vspace{-6pt}
\end{figure}

Using the numerical boson star backgrounds, we evaluate 
$\delta S_\text{half}$ as a function of mass $M$, and compare it with its counterpart in the Schwarzschild-AdS black hole, as shown in Fig.~\ref{S(M)}. 
Across the full mass range, both ground-state and excited-state boson stars exhibit substantially lower entanglement than the black holes of the same mass, which is a subthermal signature, providing evidence for scarred, non-thermal behavior, as the AdS$_3$/CFT$_2$ calculation in \cite{Liska:2022vrd}. In addition, ground-state boson stars exhibit lower entanglement than the excited solutions, showing stronger integrability as in the spectral analysis.

Even more intriguingly, three critical points emerge successively in the unstable upper branch of each 
boson star curve, associated with the maximal entropy $\delta S_{half}$, the transition between RT surface configurations, and the maximal boson-star mass $M$, respectively. Their correlated appearance suggests additional structures underlying the scarred phase. 




\vspace{.2cm}
{\bf \textit{ 
Revivals in Krylov complexity.--}} 
Krylov complexity measures the spread of an operator or state in its Krylov subspace, serving as a sharp diagnostic for quantum chaos \cite{Parker:2018yvk, Nandy:2024evd, Rabinovici:2025otw, Baiguera:2025dkc}. As a dynamical probe complementary to the spectral and entanglement diagnostics, we study the Krylov complexity of the quantum state excited by a local scalar operator, whose holographic growth has been proposed to be dual to the proper radial momentum of an infalling massive particle in the bulk \cite{Caputa:2024sux}. We use this formulation to probe the dynamical revivals in boson stars to further support the 
identification 
of a quantum scarred state.  

For the static boson star background \eqref{background fields}, we consider a massive probe particle with mass $m$ following a radial timelike geodesic. The worldline Lagrangian is
\begin{equation}
    \mathcal{L} = \frac{1}{2}g_{\mu\nu} \dot{x}^\mu\dot{x}^\nu = \frac{1}{2}\Big(-A(r)^2N(r)\dot{t}^2 + \frac{\dot{r}^2}{N(r)}\Big),
\end{equation}
where the affine parameter is chosen as proper time $\tau$. 
The radial motion and proper radial momentum are 
\begin{equation}
\begin{split}
\begin{aligned}
    \frac{dr}{dt} &= \frac{\dot{r}}{\dot{t}} = -A(r)N(r)\sqrt{1-\frac{A(r)^2N(r)}{E^2}}\,,\\
    P_\rho &= m \frac{d\rho}{d\tau} = \frac{m}{\sqrt{N(r)}}\dot{r} = -m\sqrt{\frac{E^2}{A(r)^2N(r)}-1}\,,
\end{aligned}
\end{split}
\end{equation}
where $E$ is the conserved energy determined by the initial condition at the UV cutoff, and $\rho$ is the proper radial distance. 
From the geodesic evolution we obtain $P_\rho(t)$, and evaluate the Krylov complexity through the proposal $K'(t) = -P_\rho(t)\,$.

\begin{figure}[h!]
  \centering
  \includegraphics[width=0.225\textwidth]{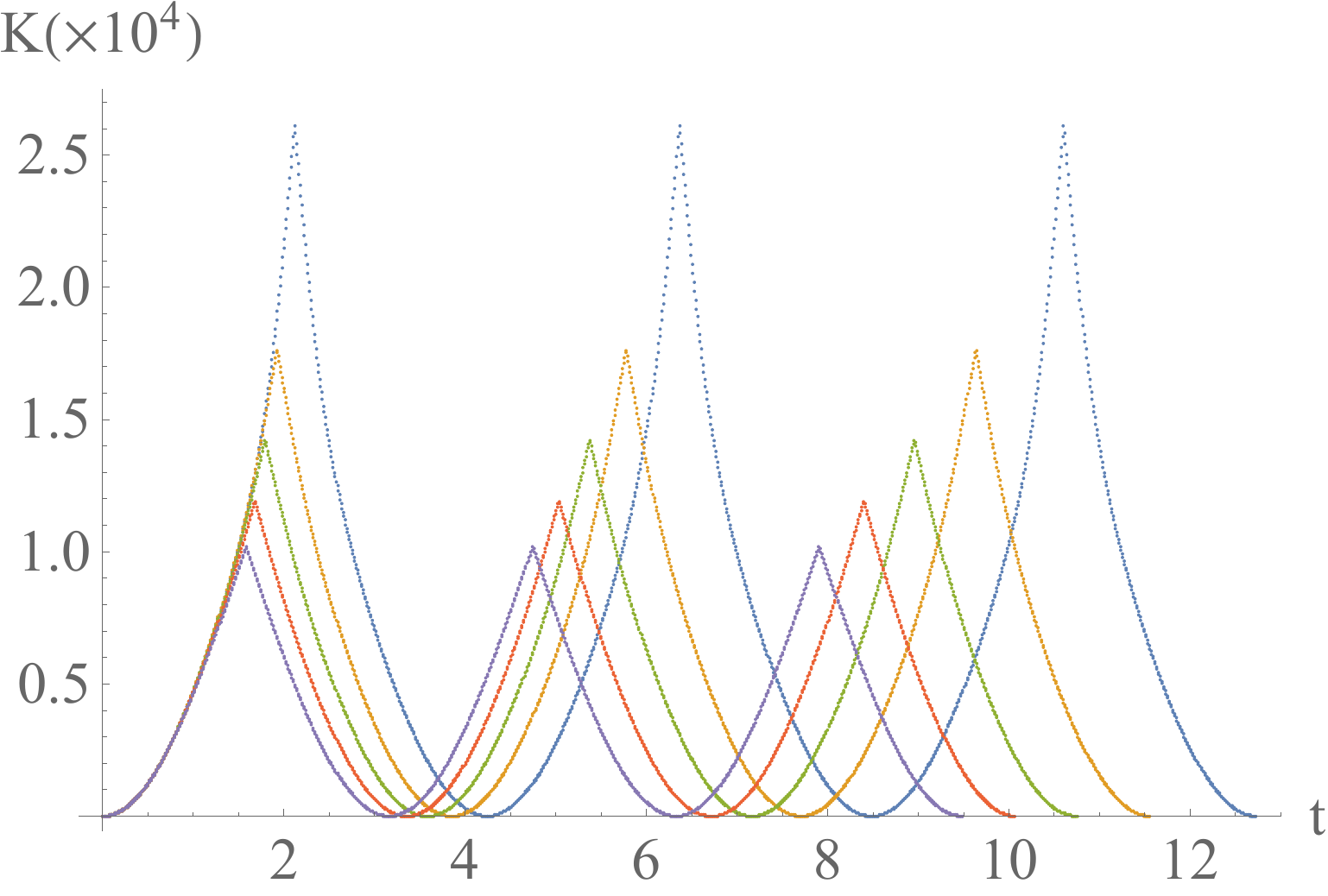}
  \includegraphics[width=0.225\textwidth]{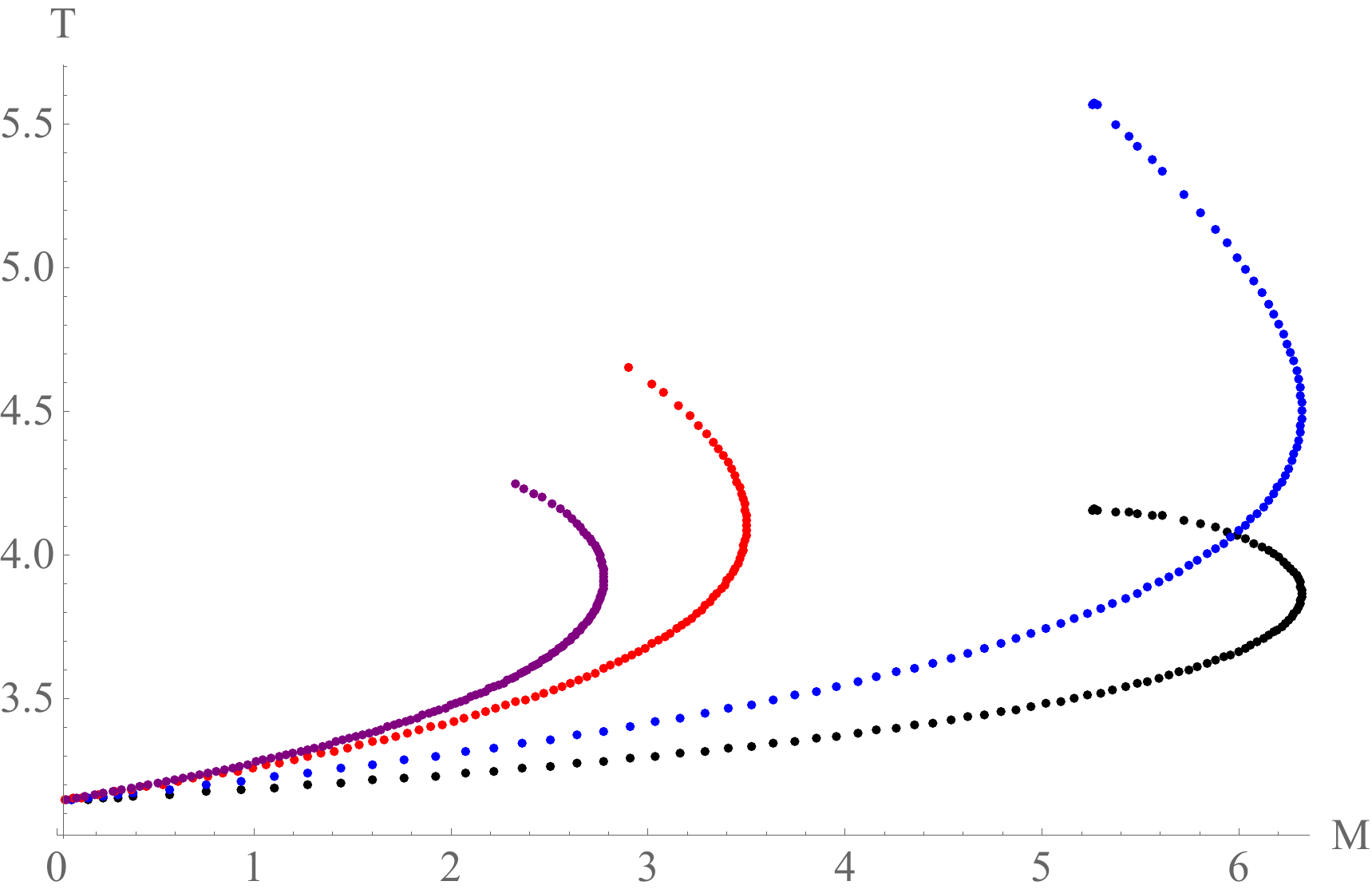}
  \caption{\label{K(t)} Holographic Krylov complexity. Left: The periodic time evolution of the complexity in the second excited stars, with $\omega = \omega_{min}\simeq 4.708$ ({\em blue}); $5$ ({\em yellow}); $5.32$ ({\em green}); $5.64$ ({\em red}); $5.96$ ({\em purple})
  . Right: The revival period as a function of the boson star mass, for the ground state star ({\em blue}), first excited star ({\em red}), and second excited star ({\em purple}), in comparison with the oscillating period $2\pi/\omega$ of the ground-state star ({\em black}).}
\vspace{-6pt}
\end{figure}


The 
typical 
time evolution of $K$ in second excited stars 
is shown in the left panel of Fig.~\ref{K(t)}. 
At early time, the growth of $P_\rho$ is linear, and therefore the growth of $K(t)$ is quadratic, consistent with the vacuum AdS behavior \cite{Caputa:2024sux}, while the growth rate increases deeper in the bulk due to the gravitation of the star. $K(t)$ reaches its maximum when the particle reaches the star center. $K(t)$ decreases
afterwards as the particle 
propagates to the opposite boundary. 
Therefore, unlike black holes, where the one-way infall leads to monotonic complexity growth, the horizonless boson star traps the particle through repeated boundary reflections. 
As a result, $K(t)$ exhibits pronounced periodic revivals. 
Note that the holographic prescription rests on the identification of the holographic radial direction with the boundary scale and operator growth. Thus, the spread of an excitation into increasingly complex Krylov components is encoded semiclassically as a particle moving deeper into the bulk spacetime. Consequently, the geometric recurrence of the particle maps to the refocusing of the boundary Krylov wave packet, signaling a scar-like revival rather than a merely kinematic bulk oscillation,
matching with the scar detection through quantum Krylov complexity \cite{Bhattacharjee:2022qjw,Nandy:2023brt,Hu:2025zvv, Caputa:2025ucl}.

The revival period $T_{rev}$ extracted from the Krylov complexity for ground-state, first excited-state, and second excited-state boson stars is shown in the right panel of Fig.~\ref{K(t)}. At fixed boson star mass, we find  $T_{rev} \geq T_g$ where $T_{g}$ is the oscillating period of the ground-state star. The inequality saturates in the vacuum limit, for which $T_{rev}=T_{g}=\pi$. This behavior differs from the quantum many-body example studied in \cite{Bhattacharjee:2022qjw} where the Krylov revival period is equal to its external driving period. We further observe that, at the same mass, $T_{rev}$ in excited stars is larger than that in ground-state stars. These relations 
suggest that the star background sets the shortest intrinsic recurrence timescale $T_g$, while the Krylov revival $T_{rev}$ captures a dressed collective recurrence of the corresponding scarred subsector, where excited stars produce stronger dressing of this recurrence. Thus, the Krylov revival provides a dynamical manifestation of the near-integrable subsectors identified in the spectral analysis.

\vspace{.2cm}
{\bf \textit{Conclusion and discussion.--}} 
We have provided evidence for AdS mini-boson stars being a holographic realization of quantum scar-like states. Through the normal-mode spectrum, we have uncovered a characteristic coexistence of chaotic spectral correlations and asymptotically integrable subsectors. We have further showed that boson stars exhibit two scar-like signatures 
beyond the spectrum: anomalously suppressed entanglement compared to thermal black holes and robust Krylov revivals associated with nonergodic dynamics. Taken together, these spectral, entanglement, and dynamical diagnostics point toward a unified picture in which the horizonless gravitational condensates realize embedded nonthermal structures within the otherwise chaotic holographic systems.

\vspace{0.2cm}
We emphasize that the scar interpretation is not based on any single feature, but on the coexistence of the correlated diagnostics. The geometries of AdS vacuum, black holes, and boson stars realize distinct regimes: (i) the AdS vacuum has a fully integrable spectrum, 
area-law entanglement scaling
, and regular Krylov revivals; (ii) the black hole displays chaotic spectral diagnostics, volume-law entanglement scaling, and rapid monotonic Krylov growth; and (iii) the boson star combines chaotic low-frequency spectral correlations with near-integrable high-frequency towers, exhibits suppressed entanglement, and supports robust Krylov revivals.

Our results suggest a new perspective on the role of coherent gravitational configurations in quantum thermalization. More broadly, they raise the possibility that quantum scars may naturally emerge in gravitational systems,
extending from many-body lattice systems to quantum gravity. It would be interesting to explore whether similar scarred structures arise more generally in rotating boson stars, fermionic stars, wormhole geometries, and nonlinear gravitational dynamics.



\vspace{0.4cm}
\section*{Acknowledgement}
\vspace{-0.1cm}
We thank Li-Ming Cao, Fu-Ming Chang, Victor Jaramillo, Hyun-Sik Jeong, Xiao-Mei Kuang, Wei-Jia Li, Jian-Xin Lu, 
Cheng Peng, Hao-Tian Sun, Hai-Qing Zhang, Shuang-Yong Zhou, and Yu-Sen Zhou for helpful discussions. 
This work was supported by the National Natural Science Foundation of China Grants No. 12375041, 12575046, 12575068, and 12247103.


\onecolumngrid
\appendix 
\clearpage
\renewcommand\thefigure{S\arabic{figure}}    
\setcounter{figure}{0} 
\renewcommand{\theequation}{S\arabic{equation}}
\setcounter{equation}{0}
\renewcommand{\thesubsection}{S\arabic{subsection}}

\section*{Supplementary Material}

\subsection{Equations of motion and boundary conditions}\label{sec S2}
In this section, we present the equations of motion for both the background and the fluctuations, along with the associated boundary conditions.

The equations of motion obtained by varying the action \eqref{eq:action} are
\begin{align}
     &G_{\mu\nu} + \Lambda g_{\mu\nu} = 8\pi G T_{\mu\nu}\,,~~~~~~ \Big(\square - \frac{\partial U}{\partial |\Phi|^2} \Big) \Phi
     =0
     \,,\nonumber\\
     &T_{\mu\nu} = -g_{\mu\nu}\big(g^{\sigma\rho}\partial_{(\sigma}\Phi^*\partial_{\rho)}\Phi + U \big) + 2\partial_{(\mu}\Phi^*\partial_{\nu)}\Phi\,.
\end{align}

Substituting the background ansatz \eqref{background fields}, we obtain the following set of coupled equations:
\begin{equation}
\begin{split}
\begin{aligned}
    0 &= r\big(-2r\phi(r)^2+N'(r)\big)+N(r)\big(1+r^2\phi'(r)^2\big)-1+r^2\Lambda+\frac{r^2\omega^2\phi(r)^2}{A(r)^2N(r)}\,,\\
    0 &= A'(r)-\frac{r\omega^2\phi(r)^2}{A(r)N(r)^2}-rA(r)\phi'(r)^2\,,\\
    0 &= N'(r)\phi'(r)+N(r)\big((\frac{2}{r}+\frac{A'(r)}{A(r)})\phi'(r)+\phi''(r)\big)+\big(2+\frac{\omega^2}{A(r)^2N(r)}\big)\phi(r)\,.
\end{aligned}
\end{split}
\end{equation}

The regularity conditions in IR, together with the asymptotically AdS boundary conditions in UV, are imposed as follows:
\begin{equation}
\begin{split}
\begin{aligned}
    N(r\rightarrow 0) &= N_0 + N_2 r^2 + \cdots,\quad N(r\rightarrow\infty) = r^2 + 1 - \frac{2GM}{r} + \cdots\,,\\
    A(r\rightarrow 0) &= A_0 + A_2 r^2 + \cdots,\quad A(r\rightarrow\infty) = 1,\\
    \phi(r\rightarrow 0) &= \Phi_0 + \Phi_2 r^2 + \cdots, \quad \phi(r\rightarrow\infty) = \frac{\phi_1}{r} + \frac{\phi_2}{r^2} + \cdots.
\end{aligned}
\end{split}
\end{equation}

\begin{figure}[h!]
  \centering
  \includegraphics[width=0.31\textwidth]{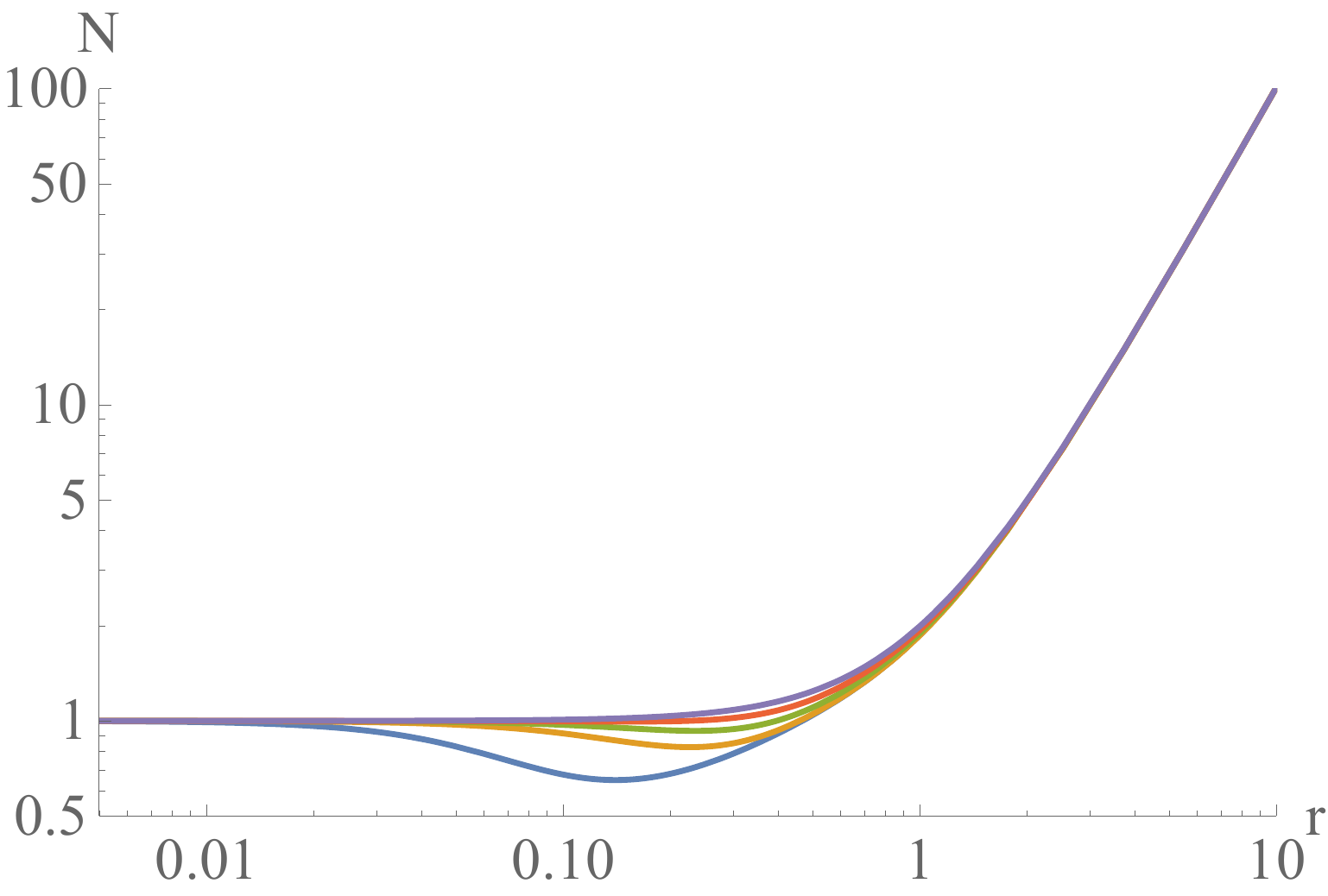}~~
  \includegraphics[width=0.31\textwidth]{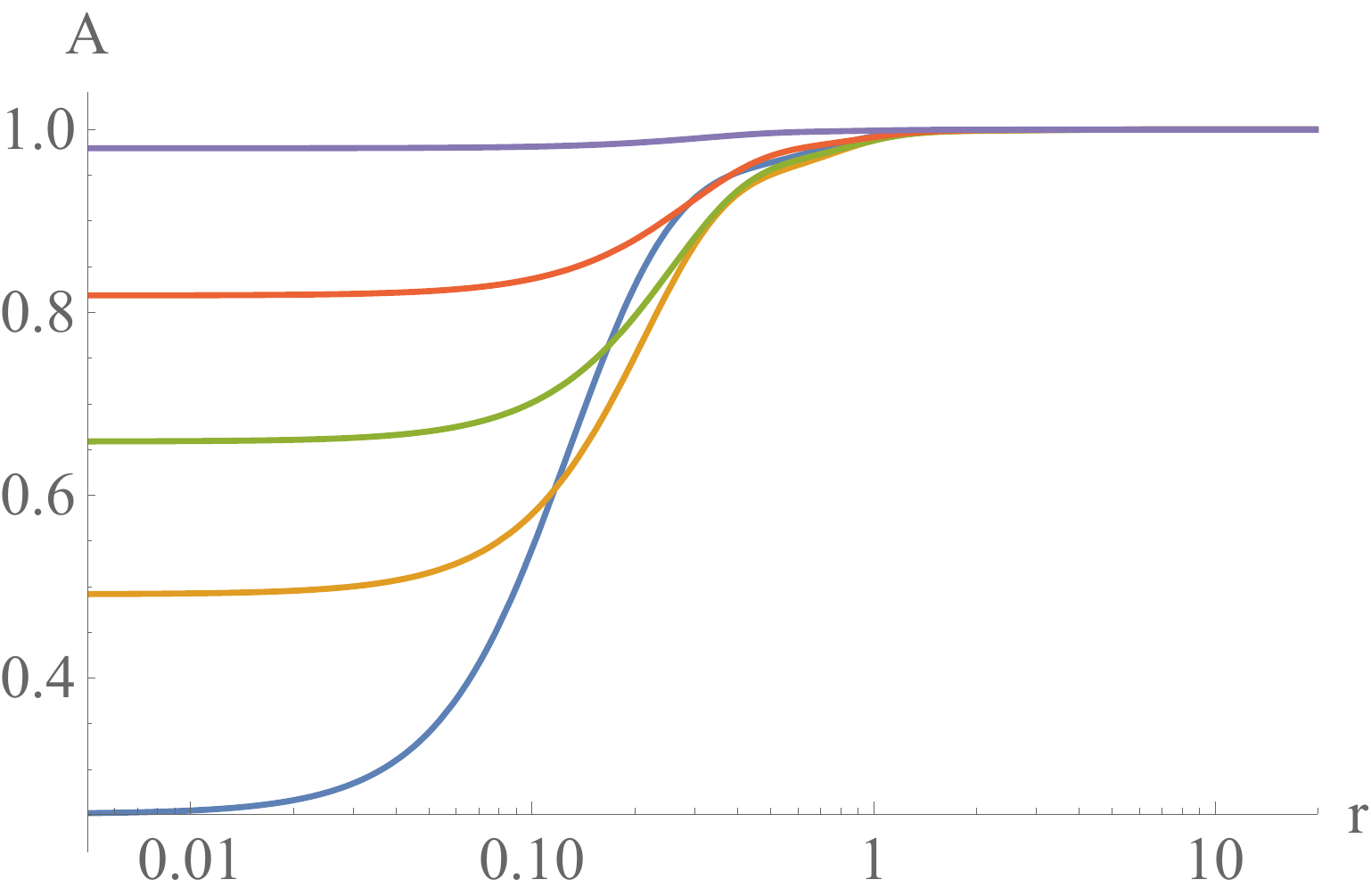}~~
  \includegraphics[width=0.31\textwidth]{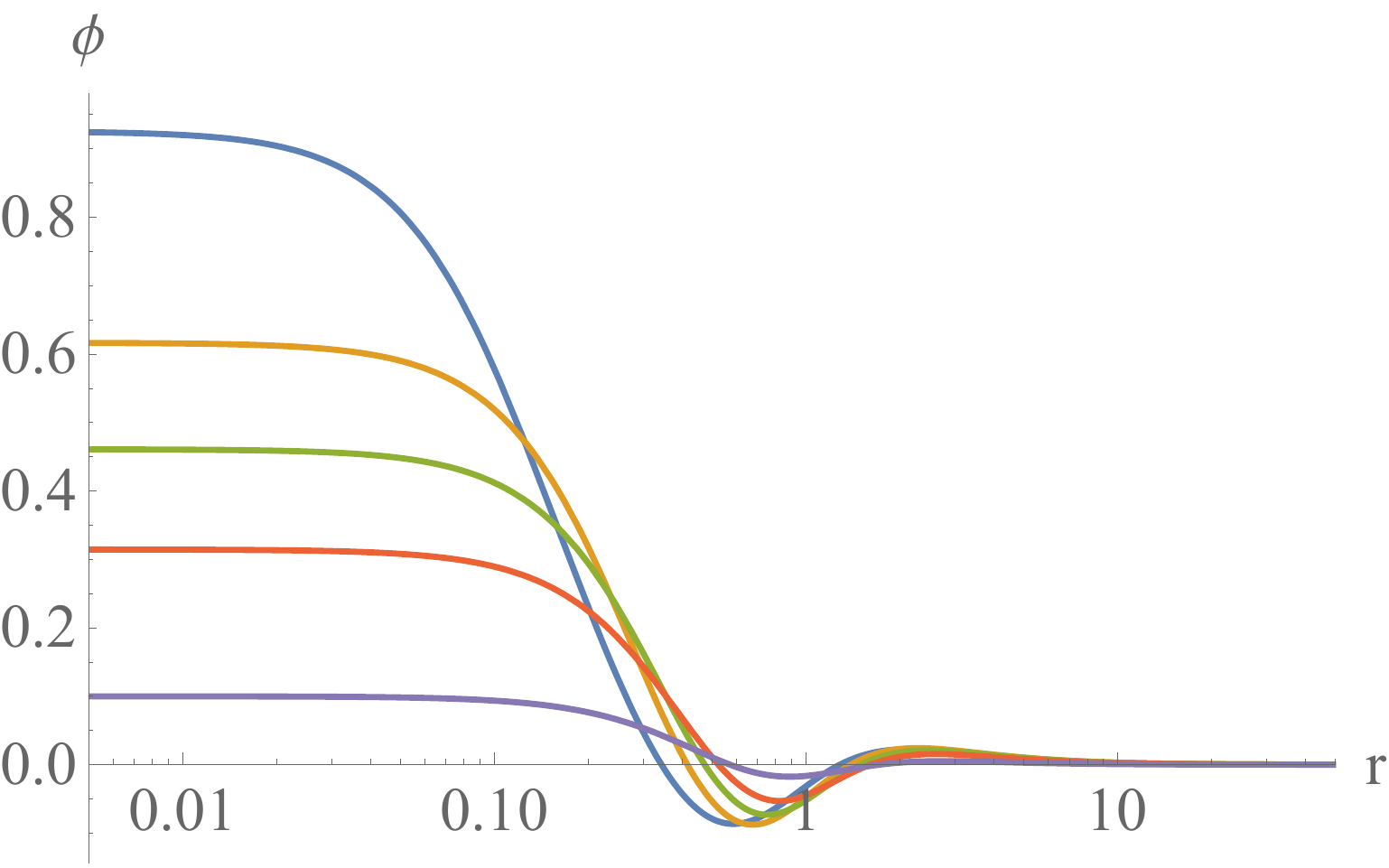}
  \caption{\label{fig background} Representative solutions of background radial fields $N(r)$ ({\em left}), $A(r)$ ({\em middle}), and $\phi(r)$ ({\em right}) for the second excited stars. Here $\omega = \omega_{min}\simeq 4.708$ ({\em blue}); $5$ ({\em yellow}); $5.32$ ({\em green}); $5.64$ ({\em red}); $5.96$ ({\em purple}). Note that the number of nodes in $\phi$ parametrizes the excited number. For the configurations of the scalar field in the ground state and first excited boson stars, one finds zero and one node, respectively.  
  }
  \label{fig:profilebg}
\vspace{-6pt}
\end{figure}

Fig. \ref{fig:profilebg} shows representative solutions for the radial metric fields $N(r)$, $A(r)$, and scalar field $\phi(r)$. The star center is located at $r=0$; the AdS boundary is at $r\to\infty$. 




The equations of motion governing the $\ell = 0$ fluctuations \eqref{eq:fluc-ell0} are
\begin{equation}
\begin{split}
\begin{aligned}
    0 &= -4 r^2 \delta\phi_-(r) \bigl(\omega (\omega - \Omega) + 2 A(r)^2 N(r)\bigr) \phi(r)
     -4 r^2 \delta\phi_+(r) \bigl(\omega (\omega + \Omega) + 2 A(r)^2 N(r)\bigr) \phi(r)+4 r^2 \omega^2 H_0(r)\phi(r)^2 \\
    &-r A(r) N(r) H_2'(r)\bigl(2 r N(r) A'(r) + A(r)(2 N(r) + r N'(r))\bigr) +r A(r) N(r) H_0'(r)\bigl(4 r N(r) A'(r) + A(r)(2 N(r) \\
    &+ 3 r N'(r))\bigr)+4 r^2 A(r)^2 N(r)^2 \delta\phi_+'(r)\phi'(r) +4 r^2 A(r)^2 N(r)^2 \delta\phi_-'(r)\phi'(r)
     +2 r^2 A(r)^2 N(r)^2 H_0''(r) -2 r H_2(r) \\
    &\Bigl(-r \Omega^2+A(r)N(r)\bigl(3 r A'(r)N'(r)+2N(r)(A'(r)+rA''(r))\bigr) +A(r)^2N(r)\bigl(2N'(r)+r(2N(r)\phi'(r)^2+N''(r))\bigr)\Bigr)\,,\\
    0 &= 2 r \delta\phi_\pm(r)\bigl((\omega\pm\Omega)^2+2A(r)^2N(r)\bigr)
     -r\omega(2\omega\pm\Omega)H_0(r)\phi(r) +2A(r)N(r)\delta\phi_\pm'(r)\bigl(rN(r)A'(r)\\
    & +A(r)(2N(r)+rN'(r))\bigr)
     -rA(r)^2N(r)^2H_2'(r)\phi'(r) +rA(r)^2N(r)^2H_0'(r)\phi'(r)
     +2rA(r)^2N(r)^2\delta\phi_\pm''(r) \\
    & +H_2(r)\Bigl(\pm r\omega\Omega\phi(r) -2A(r)N(r)\bigl(rN(r)A'(r)\phi'(r)
     +A(r)(2N(r)\phi'(r)+rN'(r)\phi'(r)+rN(r)\phi''(r))\bigr)\Bigr)\,,\\
     0 &= -\Omega H_2(r) -r\omega\phi(r)\delta\phi_+'(r) +r\omega\phi(r)\delta\phi_-'(r)
     +r(\omega+\Omega)\delta\phi_+(r)\phi'(r) +r(-\omega+\Omega)\delta\phi_-(r)\phi'(r)\,,\\
     0 &= -r^2\delta\phi_+(r)\bigl(\omega(\omega+\Omega)-2A(r)^2N(r)\bigr)\phi(r)
     +r^2\delta\phi_-(r)\bigl(\omega(-\omega+\Omega)+2A(r)^2N(r)\bigr)\phi(r) +rA(r)^2N(r)^2H_2'(r)\\
    &-r^2A(r)^2N(r)^2\delta\phi_+'(r)\phi'(r)
     -r^2A(r)^2N(r)^2\delta\phi_-'(r)\phi'(r) +A(r)^2H_2(r)N(r)\bigl(N(r)+rN'(r)+r^2N(r)\phi'(r)^2\bigr)\\
    &-A(r)^2H_0(r)N(r)\bigl(-1-3r^2+N(r)-2r^2\phi(r)^2+rN'(r)+r^2N(r)\phi'(r)^2\bigr)\,,\\
    0 &= -r^2\delta\phi_-(r)\bigl(\omega(\omega-\Omega)+2A(r)^2N(r)\bigr)\phi(r)
     -r^2\delta\phi_+(r)\bigl(\omega(\omega+\Omega)+2A(r)^2N(r)\bigr)\phi(r)  +r^2\omega^2H_0(r)\phi(r)^2\\
    & +H_2(r)\bigl(-r^2\omega^2\phi(r)^2-A(r)^2N(r)(1+3r^2+2r^2\phi(r)^2)\bigr)+rA(r)^2N(r)^2H_0'(r) \\
    & -r^2A(r)^2N(r)^2\delta\phi_+'(r)\phi'(r)-r^2A(r)^2N(r)^2\delta\phi_-'(r)\phi'(r)\,.
\end{aligned}
\end{split}
\end{equation}
{\color{black} There are three dynamical second-order ODEs for the fields $H_0(r)$ and $\delta\phi_{\pm}(r)$, together with three first-order constraint equations that determine $H_2(r)$. We have verified that all of these constraint equations are satisfied by our numerical solutions.}

The equations of motion governing the $\ell = 2$ fluctuations \eqref{eq:fluc-ell2} are
\begin{equation}
\begin{split}
\begin{aligned}
    0 &= -(-2+l+l^2)A(r)^2 K(r) N(r) +2r^2 \delta\phi_-(r)\bigl(\omega(\omega-\Omega)-2A(r)^2N(r)\bigr)\phi(r) \\
    & +2r^2 \delta\phi_+(r)\bigl(\omega(\omega+\Omega)-2A(r)^2N(r)\bigr)\phi(r)
    -2rA(r)^2N(r)^2 H_2'(r) +2r^2A(r)^2N(r)^2K''(r) \\
    & +rA(r)^2N(r)K'(r)\bigl(6N(r)+rN'(r)\bigr)
    +2r^2A(r)^2N(r)^2\delta\phi_+'(r)\phi'(r)
    +2r^2A(r)^2N(r)^2\delta\phi_-'(r)\phi'(r) \\
    & -A(r)^2H_2(r)N(r)\Bigl(-2+l+l^2-6r^2-4r^2\phi(r)^2+4rN'(r)+4N(r)\bigl(1+r^2\phi'(r)^2\bigr)\Bigr)\,, \\
    0 &= 2r^2\delta\phi_-(r)\bigl(3\omega(\omega-\Omega)+2A(r)^2N(r)\bigr)\phi(r)
    +2r^2\delta\phi_+(r)\bigl(3\omega(\omega+\Omega)+2A(r)^2N(r)\bigr)\phi(r)+4r^3\Omega^2N(r)H_1'(r) \\
    & +2r^2\Omega^2H_1(r)\bigl(4N(r)+rN'(r)\bigr) -rA(r)N(r)K'(r)\bigl(2rN(r)A'(r)+A(r)\bigl(-2N(r)+rN'(r)\bigr)\bigr) \\
    & +2rA(r)N(r)H_2'(r)\bigl(3rN(r)A'(r)+A(r)\bigl(N(r)+2rN'(r)\bigr)\bigr)  -2r^2A(r)^2N(r)^2\delta\phi_+'(r)\phi'(r) +2r^2A(r)^2N(r)^2H_2''(r)\\
    & +H_2(r)\Bigl(-2r^2\bigl(\Omega^2-2\omega^2\phi(r)^2\bigr) +2rA(r)N(r)\bigl(3rA'(r)N'(r)+2N(r)(A'(r)+rA''(r))\bigr)\\
    & -A(r)^2N(r)\bigl(-2+l+l^2-6r^2+4N(r)-4r^2\phi(r)^2-2r^2N''(r)\bigr)\Bigr) -2r^2A(r)^2N(r)^2\delta\phi_-'(r)\phi'(r)\\
    & +K(r)\Bigl(-2r^2\bigl(\Omega^2-2\omega^2\phi(r)^2\bigr)
    -2rA(r)N(r)\bigl(3rA'(r)N'(r)+2N(r)(A'(r)+rA''(r))\bigr) \\
    & -A(r)^2N(r)\bigl(-2+l+l^2-12r^2- 8r^2\phi(r)^2+4rN'(r)+4r^2N(r)\phi'(r)^2+2r^2N''(r)\bigr)\Bigr)\,,
\end{aligned}
\end{split}
\end{equation}
and
\begin{equation}
\begin{split}
\begin{aligned}    
    0 &= \delta\phi_\pm(r)\Bigl(r^2(\omega\pm\Omega)^2A(r)-\bigl(l+l^2-2r^2\bigr)A(r)^3N(r)\Bigr)
    \pm r^2\omega\Omega A(r)K(r)\phi(r) \mp r^3\omega\Omega A(r)N(r)\phi(r)H_1'(r)\\
    & +rA(r)^2N(r)\delta\phi_\pm'(r)\bigl(rN(r)A'(r)+A(r)(2N(r)+rN'(r))\bigr)-r^2A(r)^3N(r)^2H_2'(r)\phi'(r) +r^2A(r)^3N(r)^2K'(r)\phi'(r)\\
    & \mp r^2\Omega H_1(r)N(r)\bigl(-r\omega\phi(r)A'(r)+A(r)(3\omega\phi(r)+r(2\omega\pm\Omega)\phi'(r))\bigr)+r^2A(r)^3N(r)^2\delta\phi_\pm''(r) \\
    & -rA(r)H_2(r)\Bigl(-r\omega(\omega\pm\Omega)\phi(r)
    +A(r)N(r)\bigl(rN(r)A'(r)\phi'(r)+A(r)(2N(r)\phi'(r)+rN'(r)\phi'(r)+rN(r)\phi''(r))\bigr)\Bigr)\,,\\
    0 &= -2\Omega A(r)^2H_2(r)N(r) +2r\Omega A(r)^2N(r)K'(r)
    -2r\omega A(r)^2N(r)\phi(r)\delta\phi_+'(r)
    +2r\omega A(r)^2N(r)\phi(r)\delta\phi_-'(r) \\
    & -\Omega A(r)K(r)\bigl(2rN(r)A'(r)+A(r)(-2N(r)+rN'(r))\bigr)  +2r(\omega+\Omega)A(r)^2\delta\phi_+(r)N(r)\phi'(r)
    +2r(-\omega+\Omega)A(r)^2\delta\phi_-(r) \\
    & N(r)\phi'(r) +H_1(r)\Bigl(2r^2\omega^2\Omega\phi(r)^2
    -\Omega A(r)^2N(r)\bigl(-2+l+l^2-6r^2-4r^2\phi(r)^2+2rN'(r)+2N(r)(1+r^2\phi'(r)^2)\bigr)\Bigr),\\
    0 &= \Omega A(r)H_2(r)+\Omega A(r)K(r) -2\omega A(r)\delta\phi_+(r)\phi(r)
    +2\omega A(r)\delta\phi_-(r)\phi(r) -r\Omega A(r)N(r)H_1'(r)\\
    & -\Omega H_1(r)\bigl(rN(r)A'(r)+A(r)(N(r)+rN'(r))\bigr)\,,\\
    0 &= r\Omega^2 H_1(r)+A(r)^2N(r)H_2'(r)-A(r)^2N(r)K'(r)+A(r)H_2(r)\bigl(2N(r)A'(r)+A(r)N'(r)\bigr) \\
    & -2A(r)^2\delta\phi_+(r)N(r)\phi'(r) -2A(r)^2\delta\phi_-(r)N(r)\phi'(r),\\
    0 &= 4r^2\Omega^2H_1(r)N(r) +K(r)\bigl(-2r^2\Omega^2+(-2+l+l^2)A(r)^2N(r)\bigr) +2r^2\delta\phi_-(r)\bigl(\omega(\omega-\Omega)+2A(r)^2N(r)\bigr)\phi(r)\\
    & +2r^2\delta\phi_+(r)\bigl(\omega(\omega+\Omega)+2A(r)^2N(r)\bigr)\phi(r)  +H_2(r)\bigl(4r^2\omega^2\phi(r)^2-A(r)^2N(r)(-2+l+l^2-6r^2-4r^2\phi(r)^2)\bigr) \\
    & +2rA(r)^2N(r)^2H_2'(r) -rA(r)N(r)K'(r)\bigl(2rN(r)A'(r)+A(r)(2N(r)+rN'(r))\bigr) \\
    & +2r^2A(r)^2N(r)^2\delta\phi_+'(r)\phi'(r) +2r^2A(r)^2N(r)^2\delta\phi_-'(r)\phi'(r)\,.
\end{aligned}
\end{split}
\end{equation}
There are four dynamical second-order ODEs for the fields $K(r)$, $H_2(r)$, and $\delta\phi_{\pm}(r)$, together with four first-order constraint equations that determine $H_1(r)$. We have verified that all of these constraint equations are satisfied by our numerical solutions.

The IR behavior of the fluctuation fields is
\begin{equation}\label{IR BC}
\begin{split}
\begin{aligned}
    \text{for all even $\ell\geq 0$:}\\
    H_i(r\rightarrow 0) &= r^\ell(H_{i,0} + H_{i,2} r^2 + \cdots)\,,\;i = 0,\;2,\quad
    \delta\phi_\pm(r\rightarrow 0) = r^\ell(\delta\Phi_{\pm,0} + \delta\Phi_{\pm,2} r^2 + \cdots)\,,\\
    \text{additionally for $\ell\geq 2$:}\\
    H_1(r\rightarrow 0) &= r^\ell(H_{1,0} + H_{1,2} r^2 + \cdots)\,,\quad
    K(r\rightarrow 0) = r^\ell(K_0 + K_2 r^2 + \cdots)\,,
\end{aligned}
\end{split}
\end{equation}
and their UV behavior is 
\begin{equation}\label{UV BC}
\begin{split}
\begin{aligned}
    \text{for all even $\ell\geq 0$:}\\
    \delta\phi_\pm(r\rightarrow\infty) &= \frac{\delta\phi_{\pm,1}}{r} + \frac{\delta\phi_{\pm,2}}{r^2} + \cdots\,,\\
    \text{additionally for $\ell = 0$:}\\
    H_0(r\rightarrow\infty) &= h_{0,0} + \frac{h_{0,3}}{r^3} + \cdots\,,\quad
    H_2(r\rightarrow\infty) = \frac{h_{2,3}}{r^3} + \cdots\,,\\
    \text{additionally for $\ell\geq 2$:}\\
    H_1(r\rightarrow\infty) &= \frac{h_{1,2}}{r^2} + \frac{h_{1,3}}{r^3} + \cdots,\quad H_2(r\rightarrow\infty) = \frac{h_{2,1}}{r} + \frac{h_{2,2}}{r^2} + \frac{h_{2,3}}{r^3} + \cdots\,,\\
    K(r\rightarrow\infty) &= k_0 + \frac{k_1}{r} + \frac{k_2}{r^2} + \frac{k_3}{r^3} + \cdots\,.
\end{aligned}
\end{split}
\end{equation}

\subsection{Normal modes and spectral distribution functions with fixed background setting}\label{sec S3}

In this section, we first present the details of the WKB analysis and then turn to the level-spacing statistics of
both the low-lying modes and the higher-$n$ modes. 

For $\ell = 0$, the independent leading-order solutions from the WKB ansatz \eqref{eq:WKB-ansatz} take the form
\begin{equation}
\label{l=0 general sol}
\begin{split}
\begin{aligned}
    &\delta\phi_\pm^{(0)}(r) = C_+ e^{\Omega S_A(r)}\exp{\Big[-\int_\epsilon^r\Big(\frac{1}{x}\mp\frac{i\omega}{A(x)N(x)}\Big) dx\Big]}  
    + C_- e^{\Omega S_B(r)}\exp{\Big[-\int_\epsilon^r\Big(\frac{1}{x}\pm\frac{i\omega}{A(x)N(x)}\Big) dx\Big]}\\
    &\;\;~~~~~~~~\stackrel{r\rightarrow\infty}{\propto} \frac{\cos{[I_1(\Omega\pm\omega)]}}{r} + \frac{(\Omega\pm\omega)\sin{[I_1(\Omega\pm\omega)]}}{r^2} + \cdots,
\end{aligned}
\end{split}
\end{equation}
where $\epsilon$ is an IR cutoff and 
\be 
I_1 = \int_\epsilon^{r_{UV}} \frac{1}{A(x)N(x)}dx
\ee denotes the radial integral from the IR to the UV cutoff. Imposing Dirichlet boundary conditions in the UV, the two branches of normal modes, corresponding to the vanishing of  two independent scalar sources, are given by
\begin{equation}\label{l=0 modes (1)}
    \mathcal{S}_{scalar}\sim\cos{[I_1(\Omega\pm\omega)]} \;\implies\; \Omega_n = \frac{n+1/2}{I_1}\pi\mp\omega.
\end{equation}
In the vacuum AdS, these $\Omega_n$ reduce to probe scalar modes with equally spaced levels, $s=2$. Note that there is no independent gravitational mode in this sector, consistent with the fact that the $\ell = 0$ gravitational perturbation in AdS is a pure gauge.

For $\ell\geq 2$, in addition to the scalar solutions \eqref{l=0 general sol} for $\delta\phi_\pm^{(0)}(r)$, the gravitational sector contributes an independent degree of freedom:
\begin{equation}
\begin{split}
\begin{aligned}
    &H_2^{(0)}(r) = \Big(C_+ e^{\Omega S_A(r)} + C_- e^{\Omega S_B(r)}\Big) \exp{\Big[\int_\epsilon^r\Big(\frac{1}{x}-\frac{2A'(x)}{A(x)}-\frac{N'(x)}{N(x)}\Big) dx\Big]},\\
    &\;\;~~~~~~~~\stackrel{r\rightarrow\infty}{\propto}\; \frac{\cos{(I_1\Omega)}}{r} + \frac{\Omega\sin{(I_1\Omega)}}{r^2} + \cdots.
\end{aligned}
\end{split}
\end{equation}
Accordingly, in addition to the scalar modes \eqref{l=0 modes (1)}, the sourceless condition yields the gravitational spectrum 
\begin{equation}\label{l>=2 modes}
    \mathcal{S}_{grav}\sim\cos{(I_1\Omega)} \;\implies\; \Omega_n = \frac{n+1/2}{I_1}\pi,
\end{equation}
which
reduces to the standard AdS spectrum in the vacuum limit. 

For generic $\Omega_n$, the spectrum must be determined numerically. The lowest 20 modes in $\ell=0$ and $\ell=2$ sectors, for $\omega=\omega_{\text{min}}\simeq 4.708$ and $\omega=5$ in the second excited star, are listed in Tables \ref{table of l=0 modes} and \ref{table of l=2 modes}, respectively. These representative data are used to analyze the spectral statistics.  

\begin{table*}[h!]
\caption{\label{table of l=0 modes}$\ell = 0$ normal modes in the second excited star}
\begin{ruledtabular}
\begin{tabular}{ccccc}
    &\multicolumn{2}{c}{$\omega=4.708$}&\multicolumn{2}{c}{$\omega=5$}\\
    $n$ & branch 1 & branch 2 & branch 1 & branch 2\\ \hline
    1  & 0 & 1.48597 & 0 & 0.66495 \\
    2  & 1.91652 & 2.91632 & 1.59953 & 2.78003 \\
    3  & 3.67138 & 5.29007 & 3.19005 & 4.55393 \\
    4  & 6.50854 & 6.85429 & 6.25274 & 6.80997 \\
    5  & 7.96205 & 8.39021 & 7.92189 & 8.38289 \\
    6  & 9.29190 & 9.90950 & 9.57700 & 9.94685 \\
    7  & 10.68614 & 11.41835 & 11.22433 & 11.54408 \\
    8  & 12.13446 & 12.92027 & 12.86693 & 13.15997 \\
    9  & 13.59784 & 14.41741 & 14.50648 & 14.78252 \\
    10 & 15.06720 & 15.91113 & 16.14395 & 16.40824 \\
    11 & 16.54042 & 17.40234 & 17.77995 & 18.03577 \\
    12 & 18.01619 & 18.89165 & 19.41488 & 19.66439 \\
    13 & 19.49357 & 20.37950 & 21.04900 & 21.29368 \\
    14 & 20.97199 & 21.86621 & 22.68251 & 22.92339 \\
    15 & 22.45112 & 23.35200 & 24.31554 & 24.55339 \\
    16 & 23.93072 & 24.83706 & 25.94819 & 26.18357 \\
    17 & 25.41066 & 26.32151 & 27.58053 & 27.81388 \\
    18 & 26.89083 & 27.80547 & 29.21261 & 29.44427 \\
    19 & 28.37117 & 29.28901 & 30.84449 & 31.07472 \\
    20 & 29.85162 & 30.77220 & 32.47619 & 32.70521 \\
\end{tabular}
\end{ruledtabular}
\end{table*}

\begin{table*}[h!]
\caption{\label{table of l=2 modes}$\ell = 2$ normal modes in the second excited star}
\begin{ruledtabular}
\begin{tabular}{ccccccc}
    &\multicolumn{3}{c}{$\omega=4.708$}&\multicolumn{3}{c}{$\omega=5$}\\
    $n$ & branch 1 & branch 2 & branch 3 & branch 1 & branch 2 & branch 3\\ \hline
    1  & 0.82363 & 0.86665 & 2.53535 & 0.53759 & 1.12395 & 2.28545 \\
    2  & 2.92854 & 4.16781 & 4.62101 & 2.90982 & 3.87501 & 4.60310 \\
    3  & 5.60451 & 6.23139 & 6.89792 & 5.36207 & 6.23562 & 6.88235 \\
    4  & 7.75382 & 8.16773 & 8.59464 & 7.82757 & 8.44046 & 8.87292 \\
    5  & 9.18981 & 9.49608 & 10.32485 & 9.40307 & 10.02068 & 10.58440 \\
    6  & 10.57158 & 10.87096 & 11.94122 & 10.98106 & 11.61373 & 12.22528 \\
    7  & 11.97699 & 12.27198 & 13.32961 & 12.56731 & 13.21491 & 13.80293 \\
    8  & 13.52757 & 13.68958 & 14.73386 & 14.16155 & 14.82162 & 15.35478 \\
    9  & 14.96664 & 15.11868 & 16.15182 & 15.76224 & 16.43230 & 16.91144 \\
    10 & 16.31637 & 16.55620 & 17.58000 & 17.36797 & 18.04596 & 18.48125 \\
    11 & 17.64284 & 18.00007 & 18.99106 & 18.97762 & 19.66192 & 20.06290 \\
    12 & 19.01784 & 19.44887 & 20.36914 & 20.59036 & 21.27972 & 21.65369 \\
    13 & 20.45996 & 20.90156 & 21.76682 & 22.20556 & 22.89900 & 23.25143 \\
    14 & 21.90774 & 22.35749 & 23.17888 & 23.82275 & 24.51951 & 24.85453 \\
    15 & 23.35970 & 23.81606 & 24.60185 & 25.44158 & 26.14104 & 26.46181 \\
    16 & 24.81506 & 25.27685 & 26.03334 & 27.06175 & 27.76345 & 28.07240 \\
    17 & 26.27324 & 26.73952 & 27.47158 & 28.68304 & 29.38659 & 29.68565 \\
    18 & 27.73379 & 28.20380 & 28.91527 & 30.30529 & 31.01037 & 31.30107 \\
    19 & 29.19635 & 29.66947 & 30.36343 & 31.92835 & 32.63469 & 32.91826 \\
    20 & 30.66063 & 31.13636 & 31.81529 & 33.55210 & 34.25955 & 34.53697 \\
\end{tabular}
\end{ruledtabular}
\end{table*}

\begin{figure}[h!]
\centering
\includegraphics[width=0.8\textwidth]{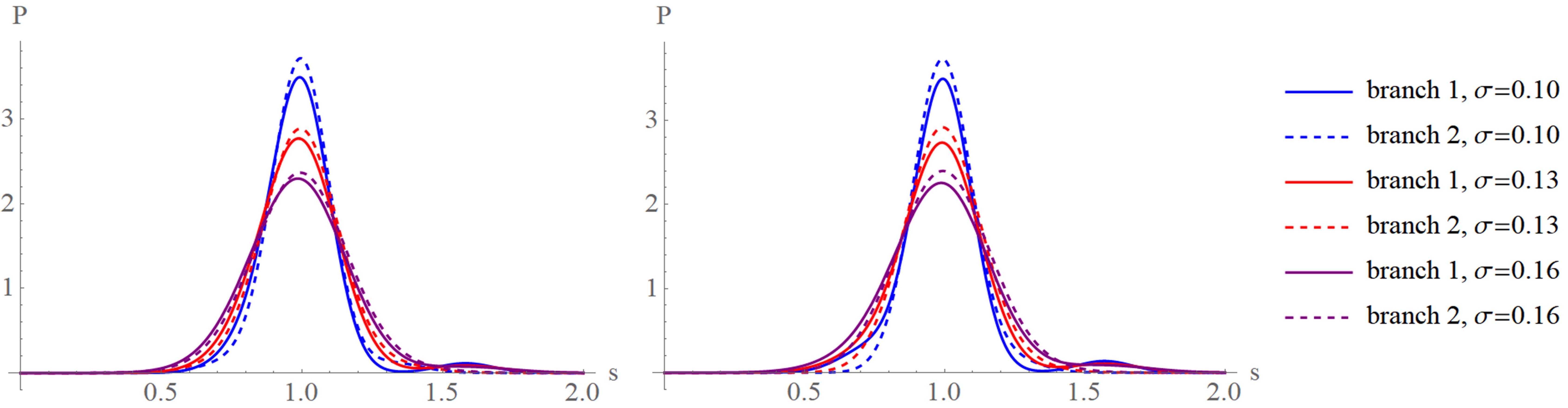}
\caption{\label{P(s)(0)} Spectral distribution functions of the WKB branches with specified variance $\sigma$ for $\ell=0$ sector in the second excited star. Left: $\omega=4.708$; right: $\omega=5$.}
\end{figure}

\begin{figure}[h!]
\centering
\includegraphics[width=0.8\textwidth]{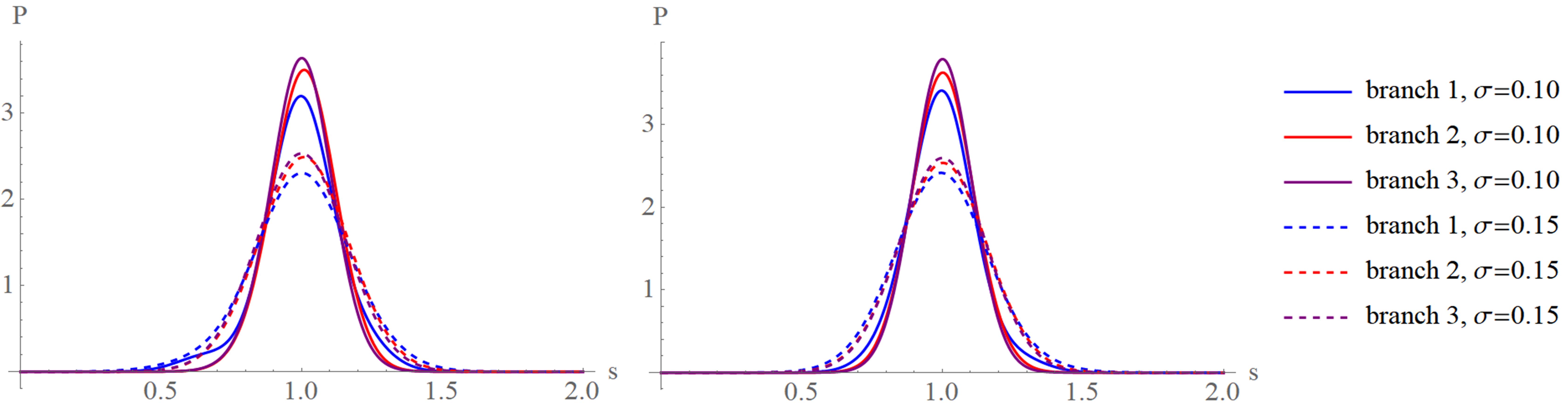}
\caption{\label{P(s)(2)} Spectral distribution functions of the WKB branches with specified variance $\sigma$ for $\ell=2$ sector in the second excited star. Left: $\omega=4.708$; right: $\omega=5$.}
\end{figure}

The spectral distribution functions $P(s)$, constructed from the first 70 unfolded levels, are shown in Figs. \ref{P(s)(0)} and \ref{P(s)(2)}. As $n$ increases, all branches of modes exhibit asymptotically equal level spacings. Consequently, $P(s)$ has a delta-like peak indicating non-Poissonian integrability. This behavior is consistent with the WKB result which predicts equally spaced energy levels for higher $n$.  
The statistical properties of the spectra in 
horizonless holographic systems were also investigated in {\em e.g.} \cite{Anegawa:2024wov, Begines:2026fnx}.

\subsection{Review of the PXP model and scar detections}\label{sec S1}

Quantum many-body scars were first systematically identified in constrained Rydberg atom arrays, whose dynamics are effectively described by the PXP model \cite{Turner:2017fxc, Bernien:2017ubn, Serbyn:2020wys, Moudgalya:2021xlu}. This model provides a minimal setting exhibiting robust non-ergodic dynamics within an otherwise chaotic spectrum, and has become a standard benchmark for studying scarred structures. Its well-established diagnostics, such as spectral statistics, entanglement entropy, and Krylov complexity, motivate the corresponding holographic probes employed in our analysis of the boson stars. 
 
The PXP model describes a chain of Rydberg atoms subject to the blockade constraint, which forbids simultaneous excitation of neighboring atoms from the ground state $\mid\downarrow\rangle$ to the excited state $\mid\uparrow\rangle$. The Hamiltonian is
\begin{equation}
    H_{\text{PXP}} = \sum_{i=1}^{L} P_{i-1} X_i P_{i+1}\,,\quad X_i = \mid\downarrow_i\rangle \langle \uparrow_i\mid + \mid\uparrow_i\rangle \langle \downarrow_i\mid,\quad P_i = \mid\downarrow_i\rangle \langle \downarrow_i\mid\,,
\end{equation}
where $X_i$ flips the atomic state at site $i$, and $P_i$ projects onto the ground state, thereby enforcing the constraint. 

The quantum dynamics are governed by the Schrödinger equation
\begin{equation}
    i \frac{d}{dt} |\psi(t)\rangle = H_{\text{PXP}} |\psi(t)\rangle\,,
\end{equation}
leading to the eigenvalue problem 
\begin{equation}
    H_{\text{PXP}} |\psi_n\rangle = E_n |\psi_n\rangle\,.
\end{equation}
The spectrum ${E_n}$ forms the basis for statistical diagnostics. A commonly used probe state is the Néel state
$|Z_2\rangle = \mid\uparrow\downarrow\uparrow\downarrow\uparrow\downarrow \cdots\rangle$, which exhibits long-lived coherent oscillations due to its large overlap with a special subset of eigenstates.

In generic chaotic quantum systems, the unfolded energy spectrum follows random matrix statistics. A standard diagnostic is the average gap ratio $\langle r\rangle$:
\begin{equation}
\label{eq:gapratio}
    s_n = E_{n+1} - E_n\,, \quad r_n = \frac{\text{min}(s_n,s_{n-1})}{\text{max}(s_n,s_{n-1})}\,, \quad \langle r\rangle = \bar{r}_n\,,
\end{equation}
where $s_n$ is the level spacing and $r_n$ is the gap ratio of neighboring eigenvalues. The ensemble average distinguishes integrable and chaotic behavior: 
\bea
\begin{split}
&&\langle r\rangle = 0.38629 \text{~for Poisson ensemble},~~ \langle r\rangle = 0.53590 \text{~for Gaussian orthogonal ensemble}, \\&&\langle r\rangle = 0.60266 \text{~for Gaussian unitary ensemble},~~ \langle r\rangle = 0.67617 \text{~for Gaussian symplectic ensemble}. 
\end{split}
\eea
In the PXP model, most eigenstates follow GOE statistics, while a small subset of anomalous eigenstates has a large overlap with the Néel state and forms a nearly equally spaced sequence. These states form the scar tower, violating the eigenstate thermalization hypothesis. 

For the entanglement diagnostics, the bipartite entanglement entropy is defined as
\begin{equation}
    S_A = - \text{Tr}(\rho_A \ln \rho_A), \quad \rho_A = \text{Tr}_{\bar{A}} |\psi_n\rangle \langle \psi_n|\,,
\end{equation}
where $\rho_A$ is the reduced density matrix. In chaotic systems, $S_A$ typically follows a volume law, $S_A \propto |A|$, whereas scarred eigenstates exhibit anomalously low entanglement, as typically observed in the PXP model \cite{Turner:2018yco, Ho:2018rum}. The time-dependent oscillations of the entanglement entropy in scarred systems were discussed in \cite{Alhambra:2019xaa, ODea:2024yqr}. In the holographic setting, the entanglement entropy is computed via the prescription in \cite{Ryu:2006bv}. 

To characterize the growth of states, we employ the Krylov formalism \cite{Baiguera:2025dkc}. For consistency, we briefly summarize the state version of the formalism. The equivalent operator version was originally proposed in \cite{Parker:2018yvk}. Starting from a normalized initial state $|\psi(0)\rangle$, one constructs an orthonormal Krylov basis ${|\phi_n\rangle}$ and Lanczos coefficients $\{a_n,b_n\}$ iteratively: 
Fix $|\phi_0\rangle=|\psi(0)\rangle,\;|\phi_{-1}\rangle=0,\;b_0=0$, and at the $n$-th step evaluate successively
\begin{equation}
\begin{split}
\begin{aligned}
    &\text{diagonal coefficient } a_n = \langle\phi_n|H|\phi_n\rangle\,,\\
    &\text{residual vector } |\tilde{\phi}_{n+1}\rangle = H|\phi_n\rangle - a_n|\phi_n\rangle - b_n|\phi_{n-1}\rangle\,,\\
    &\text{off-diagonal coefficient } b_{n+1} = \sqrt{\langle\tilde{\phi}_{n+1}|\tilde{\phi}_{n+1}\rangle}\,,\\
    &\text{basis state } |\phi_{n+1}\rangle = |\tilde{\phi}_{n+1}\rangle/b_{n+1}\,.
\end{aligned}
\end{split}
\end{equation}
The time-evolved state can be expanded in the Krylov subspace
\begin{equation}
    |\psi(t)\rangle = e^{-iHt} |\psi(0)\rangle = \sum_{n=0}^\infty \phi_n(t)|\phi_n\rangle\,,
\end{equation}
where the time-dependent coefficients up to a large cutoff $N$ are determined by
\begin{equation}
\begin{split}
\begin{aligned}
    i\dot{\phi}_n(t) &= b_{n+1}\phi_{n+1}(t) + a_n\phi_n(t) + b_n \phi_{n-1}(t)\,,\\
    \phi_0(0) &= 1\,,\;~~\;\phi_{n\geq 1}(0) = 0\,,\;~~\;\phi_{N+1}(t) = 0\,,
\end{aligned}
\end{split}
\end{equation}
where the dot denotes the derivative of $t$. The Krylov complexity is then defined as
\begin{equation}
    K(t) = \sum_{n=0}^\infty n |\phi_n(t)|^2\,,
\end{equation}
which quantifies the spread of the state in its Krylov subspace. In the PXP model, the Krylov complexity of the Néel state exhibits periodic revivals rather than unbounded growth, 
{\em i.e.}, $K(t+T)=K(t)$, providing a clear signature of scarred dynamics. In the ``Revivals in Krylov complexity" section, we adopt the corresponding holographic construction proposed in \cite{Caputa:2024sux}.

\end{document}